\documentclass[11pt,preprint]{aastex}
\slugcomment{Submitted to the Astrophysical Journal}
\shorttitle{Photoionization of Galactic Halo Gas by SNRs}
\shortauthors{Slavin, McKee \& Hollenbach}
\begin{document}
\newcommand{\acg}{{\cal A}_{\rm Cg}}
\newcommand{\Ha}{H$\alpha$}
\newcommand{\NCIIe}{$N$(\ion{C}{2}$^*$)}
\newcommand{\CIIe}{\ion{C}{2}$^*$}
\newcommand{\NHI}{$N$(\ion{H}{1})}
\newcommand{\NHII}{$N$(\ion{H}{2})}
\newcommand{\NHeII}{$N$(\ion{He}{2})}
\newcommand{\Hy}{{\rm H}}
\newcommand{\Ho}{{{\rm H}^0}}
\newcommand{\Hp}{{{\rm H}^+}}
\newcommand{\He}{{{\rm He}}}
\newcommand{\Heo}{{{\rm He}^0}}
\newcommand{\Hep}{{{\rm He}^+}}
\newcommand{\Hepp}{{{\rm He}^{++}}}
\newcommand{\Cps}{{{\rm C}^{+*}}}
\newcommand{\Su}{{{\rm S}}}
\newcommand{\Sp}{{{\rm S}^+}}
\newcommand{\Spp}{{{\rm S}^{++}}}
\newcommand{\NHo}{N_{\Ho}}
\newcommand{\NHp}{N_{\Hp}}
\newcommand{\NHeo}{N_{\Heo}}
\newcommand{\NHep}{N_{\Hep}}
\newcommand{\NHepp}{N_{\Hepp}}
\newcommand{\NHX}{N_{\Hy,X}}
\newcommand{\HI}{\ion{H}{1}}
\newcommand{\mHI}{\mathrm{H}\;\mathrm{\textsc{i}}}
\newcommand{\mHII}{\mathrm{H}\;\mathrm{\textsc{ii}}}
\newcommand{\mHeI}{\mathrm{He}\;\mathrm{\textsc{i}}}
\newcommand{\mHeII}{\mathrm{He}\;\mathrm{\textsc{ii}}}
\newcommand{\HeI}{\ion{He}{1}}
\newcommand{\HeII}{\ion{He}{2}}
\newcommand{\NSII}{$N$(\ion{S}{2})}
\newcommand{\NSIII}{$N$(\ion{S}{3})}
\newcommand{\NHop}{N_{\Ho\perp}}
\newcommand{\NHoc}{N_{\Ho c}}
\newcommand{\cc}{cm$^{-3}$}
\newcommand{\sqc}{cm$^{-2}$}
\newcommand{\ccK}{{cm$^{-3}\,$K}}
\newcommand{\tcn}{\tau_{c\nu}}
\newcommand{\caln}{{\cal N}}
\newcommand{\meps}{\langle\epsilon_{\nu}\rangle}
\newcommand{\ho}{H$^0$}
\newcommand{\jsnb}{\bar J_\nu}
\newcommand\beq{\begin{equation}}
\newcommand\eeq{\end{equation}}
\newcommand\beqa{\begin{eqnarray}}
\newcommand\eeqa{\end{eqnarray}}
\newcommand\e{$^{-1}$}
\newcommand\ee{$^{-2}$}
\newcommand\eee{$^{-3}$}
\newcommand\ton{{\tau_{0\nu}}}
\newcommand\alt{\alpha^{(2)}}
\newcommand\calng{{\cal N}_\gamma}
\newcommand\empc{EM_{\rm pc}}
\newcommand\emppc{EM_{\perp\rm pc}}
\newcommand\epn{\epsilon_\nu}
\newcommand\epna{\langle\epsilon_{\nu A}\rangle}
\newcommand\epias{\langle\epsilon_{i A}^*\rangle}
\newcommand\Herom{{\rm He^0}}
\newcommand\ephas{\langle\epsilon_{\Hy A}^*\rangle}
\newcommand\epheas{\langle\epsilon_{\Heo A}^*\rangle}
\newcommand\fnm{F_{\nu-}}
\newcommand\fnp{F_{\nu+}}
\newcommand\mha{{\rm H}\alpha}
\newcommand\jnb{\bar J_\nu}
\newcommand\nib{\bar\nu_i}
\newcommand\nlos{{\cal N}_{\rm los}}
\newcommand\phig{\phi(>\nu_i)}
\newcommand\R{{\rm R}}

\title{Photoionization of Galactic Halo Gas by Old Supernova Remnants}

\author{Jonathan D. Slavin\altaffilmark{1}}
\affil{Department of Astronomy, University of California,
Berkeley, CA 94720}

\author{Christopher F. McKee}
\affil{Departments of Physics and Astronomy, University of California,
Berkeley, CA 94720}

\and

\author{David J. Hollenbach}
\affil{NASA/Ames Research Center, MS 245-3, Moffett Field, CA 94035}

\altaffiltext{1}{Postal address: NASA/Ames Research Center, MS 245-3, 
Moffett Field, CA 94035}

\begin{abstract}
We present new calculations on the contribution from cooling hot gas to the
photoionization of warm ionized gas in the Galaxy.  We show that hot gas in
cooling supernova remnants (SNRs) is an important source of photoionization,
particularly for gas in the halo.  We find that in many regions at high
latitude this source is adequate to account for the observed ionization so
there is no need to find ways to transport stellar photons from the disk. The
flux from cooling SNRs sets a floor on the ionization along any line of sight.
Our model flux is also shown to be consistent with the diffuse soft X-ray 
background and with soft X-ray observations of external galaxies.  

We consider the ionization of the clouds observed towards the halo star
HD~93521, for which there are no O stars close to the line of sight.  Along
this line of sight, two groups of clouds (densities $\sim 0.3$--1 cm$^{-3}$
and temperatures $\sim 7000$ K) are observed at $\sim 0$ km s$^{-1}$
(``slow'') and $\sim -50$ km s$^{-1}$ (``fast'').  We show that the observed
ionization can be explained successfully by our model EUV/soft X-ray flux from
cooling hot gas.  In particular, we can match the \Ha\ intensity, the
$\Spp/\Sp$ ratio, and the $\Cps$ column.  Our models show that it is possible
to account for the observed ionization without invoking exotic ionization
mechanisms such as decaying neutrinos \citep{Sc90}.  Our value for X-ray
opacity along this line of sight is somewhat larger than the average for the
halo found by \citet{AB99}, but we do not regard the difference as
significant.  From observations of the ratios of columns of $\Cps$ and either
$\Sp$ or $\Ho$, we are able to estimate the thermal pressure in the clouds.
The slow clouds require high ($\sim 10^4$ cm$^{-3}$ K) thermal pressures to
match the $N_\Cps/N_\Sp$ ratio.  Additional heating sources are required for
the slow clouds to maintain their $\sim 7000$ K temperatures at these
pressures, as found by \citet{RHT99}.

\end{abstract}


\section{Introduction}

The nature of diffuse warm ionized gas (or warm ionized medium, WIM) in the
interstellar medium has been puzzling since its discovery.  In a series of
articles, \citet[and Reynolds \& Tufte 1995]{R84,R85a,R85b,R88,R89a,R89b} has
pointed out several unusual properties of this gas including: high
[\ion{S}{2}] $\lambda$6716/\Ha\ ratios; low [\ion{O}{1}] $\lambda$6300/\Ha ,
[\ion{O}{3}] $\lambda$5007/\Ha\ and \ion{He}{1} $\lambda$5876/\Ha\ ratios; high
scale height; and large power requirement for its ionization.

Among the characteristics of the WIM that are the most difficult to make sense
of are its power requirement, its large scale height and its relative
smoothness. The power required to maintain the ionization has been estimated
by \citet{R84} to be $9\times10^{-5}\,$ ergs s$^{-1}$ cm$^{-2}$, putting
severe constraints on the source of ionizing photons.  Only OB stars and
supernovae appear to fulfill this power requirement.  Citing the high
efficiency required for conversion of explosion energy into photoionization,
\citet{R84} favored early type stars as the primary source for maintaining
the WIM.  There are two principal difficulties with this source,
however.  First, several lines of sight in the plane apparently do not pass
sufficiently close to any O or B stars \citep{R90}.  Equivalently, the
relative smoothness of the \Ha\ background appears inconsistent with the very
clumpy distribution of early type stars.  Second, the large scale height
observed for the emission, $\sim 1\,$kpc, is much higher than the stellar
scale height.  This problem has been addressed by \citet{MC93} and by
\citet{DS94} by modeling the escape of Lyman continuum photons from the disk,
particularly from OB associations.  For some cases, however, even the
existence of large ``HII chimneys'' cannot explain the ionization at high
latitudes.  A prime example of this is the line of sight towards HD~93521 that
we discuss in detail below.

Cooling hot gas avoids some of these difficulties.  First, although supernovae
should have a scale height essentially identical to OB stars, hot gas is
buoyant in the galactic gravitational field and rises to much larger scale
heights.  This is directly observed in the large scale heights of the highly
ionized species C$^{+3}$, N$^{+4}$ and Si$^{+3}$ \citep[][and references
therein]{SSL97}.  Second, hot gas by its very nature is a more
smoothly distributed source of ionizing radiation than stars.  While the
effect of a single star or star cluster drops off with the square of the
distance, the flux from hot gas, even distributed non-uniformly in regions of
high emissivity, depends on the line of sight emission measure and thus tends
to be much smoother.  Nevertheless, the high efficiency required for
conversion of SN energy into photoionizations remains.  We do not argue in
this paper that old supernova remnants are the sole or even primary source of
the ionization responsible for the entire diffuse galactic \Ha\ background.
Rather we show that cooling hot gas sets a floor on the ionization level for
diffuse gas in the ISM in general and the galactic halo in particular.
Moreover, for at least the line of sight towards HD~93521 and most probably
for many high galactic latitude sight lines, cooling hot gas \emph{is} the
dominant source of the ionization.  Thus we find that in many cases there is
no need to devise a means of transporting photons from early type stars in the
disk to the galactic halo.

The ionizing flux produced by cooling hot gas cannot be measured directly
because the photons responsible for most of the ionization lie in the 13.6 --
50 eV range, a part of the EUV for which the diffuse background has not yet
been definitively detected \citep[see][]{VS98}.  We are therefore left to
theoretical modeling to estimate the ionizing flux.  Our procedure, detailed
below, is to calculate a space and time averaged emissivity due to supernova
remnants expanding in the ISM.  The mean emissivity (along with an estimate of
the fraction of ionizing photons that escape from the disk) gives us a
prediction for the emission measure generated by photoionization due to hot
gas emission. Using the emissivity and a simple model for the mean opacity in
a cloudy medium interspersed with hot gas, we derive the mean intensity
incident on a cloud face.  [It should be noted that we use the term ``cloud''
to refer to any cold or warm ($T\la 10^4$~K) region of gas that is embedded in
much lower density ($n\sim 5\times10^{-3}$ \cc), hot ionized medium (HIM).] We
are then able to calculate such observables as the C$^{+*}$ (C$^+$ in the
excited fine structure level), S$^+$, and S$^{++}$ column densities per cloud
as well as a typical value for the galactic soft X-ray background and the
X-ray surface brightness of the Galaxy as viewed from the outside.  As we show
below, our model is remarkably successful in explaining a variety of
observations.  Thus, while our model is clearly too simple to explain all the
detailed observations of the soft X-ray background and ionization in the halo,
its success provides evidence that cooling hot gas is an important source of
ionization for the WIM.

In \S 2 below we describe current evidence on emission from old SNRs and give
details about our model calculations of that emission.  In \S 3 we use our
model to predict the emission measure and \Ha\ intensity generated by
ionization due to hot gas emission and compare with observations.  In \S 4 we
look at other ionization predictions from the model and compare with
observations towards the line of sight towards HD~93521.  We discuss how
observations of \ion{S}{2}, \ion{S}{3}, and \CIIe\ absorption lines constrain
the morphology of the clouds and their thermal pressure, whereas [\ion{N}{2}],
[\ion{S}{2}] and [\ion{O}{1}] emission lines constrain the temperature.  In \S
5 we describe how the ionization of helium in our model affects the X-ray
opacity of the halo. In \S 6 we discuss the uncertainties in the model
spectrum, the need for additional heating in the slow clouds, and the decaying
neutrino model for the ionization proposed by \citet{Sc90}.  Finally, we
summarize  our conclusions in \S 7.

\section{The Soft X-ray Background from Old Supernova Remnants}

\subsection{Observations of the Local Background}

The direct evidence for soft X-ray emission from hot gas comes primarily from
proportional counter observations of the diffuse background by the Wisconsin
group \citep[e.g.,][]{MBSK83} and by ROSAT \citep[e.g.,][]{Sea97}.  The most
important of these for the issue of ionization are the lower energy bands: the
B and C bands ($\sim 150$ and 250 eV) of the Wisconsin instruments and the R1
and R2 bands ($\sim 200$ and 250 eV) defined for ROSAT.  The all-sky maps
produced by both sets of observations are essentially consistent with each
other when allowances are made for the difference in filters, sensitivity and
angular resolution \citep{Sea95}.

In the broadest terms, the observations show the whole sky to be glowing with
soft X-ray emission with a general trend of greater brightness towards the
poles.  While there has been much discussion of the proper interpretation of
the results \citep[e.g.,][]{C98,M98}, there is now a general consensus that
most of the observed emission comes from a cavity in the neutral hydrogen that
surrounds the Sun and is filled with hot ($T \approx 10^6\,$K), very low
density gas.  At high latitude the 1/4 keV (C band) background also has
substantial contributions from more distant emission, in some directions
dominating the local component.  In the galactic plane this seems not to be
the case, though the presence of low column density clouds, seen by optical
and EUV absorption, has yet to be fully taken into account in the modeling of
the observed flux \citep[see][]{S98}.

\subsection{Models for the Soft X-ray Background}

Several studies have been carried out to explain the origins of the soft X-ray
emission.  \citet{MO77} put forward a comprehensive model for the
diffuse ISM in which supernova remnants sweep the warm and cold gas into
shells and isolated clouds, leaving most of the medium filled with hot gas.
The observed local emission is seen as being somewhat greater than in the
typical ISM due to our being in an old supernova remnant (their estimate
corresponds to an age of about $3\times 10^5$ yr).  Cox \& collaborators
\citep{CA82,E86,EC93,SC98} have explored a number of models aimed specifically
at explaining the SXRB as the result of the Sun's position inside a supernova
remnant (possibly due to several explosions).  Our model that we present below
is not aimed specifically at explaining the observed SXRB and is, in this
sense, more like the McKee \& Ostriker model.  We do not attempt, however, to
model the medium as a whole and our model of the emissivity does not rest on
any particular model for the morphology of the ISM.  On the other hand, our
model for the flux incident on clouds does require assumptions about the
distribution of opacity in the medium and we use this fact below to draw
conclusions on typical cloud (i.e., WNM and WIM) sizes.  In addition, the data
for the line of sight towards HD~93521 gives us information on the typical
pressures in clouds that we can use to set limits to the filling factor of
warm clouds in this direction.  We find, in accordance with \citet{R91} and
\citet{SF93}, that the warm clouds have a fairly small filling factor,
$\lesssim 10$\% (see \S \ref{sect:modres} below).

A model for the mean emission from a population of evolving SNRs has
previously been calculated by \citet{CM86} for the purpose of constraining the
properties of SNRs in the galaxy M101.  They carried out analytical
calculations of the distribution over temperature of the surface averaged
emission measure. A set of numerical calculations similar to the ones we
present here was performed by \citet{M94} again with observations of external
galaxies in mind.  The aim of our calculations is somewhat different, being
focussed on the ionizing properties of the x-ray/EUV emission.  At the time of
publication of \citet{CM86}, only upper limits to the soft X-ray emission were
available.  With the advent of \emph{ROSAT}, the X-ray emission has been
detected for M101 and a number of other galaxies.  As we discuss below, the
fraction of soft X-ray emission that escapes the galaxy provides one of the
tests of our model.  In addition, we examine in more detail the spectrum
produced by the cooling, supernova-shocked gas and its effect on the
ionization structure of the WIM and the WNM.

\subsection{The Hot Gas Emission Model}

The primary purpose of this paper is to explore the effects of photoionization
by cooling hot gas in the galaxy.  As a consequence, we have focussed on the
calculation of the emission and have made very simple assumptions regarding
the ISM.  To calculate the mean emissivity, we assume independent evolution of
SNRs expanding in a uniform density medium.  The mean emissivity of the hot
gas is then
\begin{equation}
\meps = S \int\! dt \int\! \epsilon_{\nu}\,dV
\label{eq:meps}
\end{equation}
where $S$ is the supernova rate per unit volume, $\epsilon_{\nu}$ is the
emissivity per unit frequency in the remnant as a function of position and
time within the remnant.  The integration is carried out over the volume of
the remnant and time evolution of the remnant.  We shall generally work with
quantities measured per unit area of the disk, so we define $S_A\equiv\int S
dz$ as the supernova rate per unit area and $\epna\equiv\int\meps dz$ as the
mean emissivity per unit area.  Normalizing to the explosion energy, 
$E_{\rm SN}$ we define:
\begin{equation}
\phi_{\nu} \equiv \frac{\meps}{S E_{\rm SN}}
=\frac{\epna}{S_AE_{\rm SN}}.
\label{eq:phinu}
\end{equation}
Thus $\phi_{\nu}$ is the fraction of the supernova power radiated per unit
frequency interval.  Equation (\ref{eq:phinu}) shows that the total supernova
power, or equivalently, the supernova power per unit disk area, is separable
from the calculation of the mean spectrum that determines the emission
characteristics of the SNRs.  This formulation brings out the fact that it is
this distribution in frequency that is critical in determining the
photoionizing properties of the cooling hot gas in SNRs.  

To carry out the calculations of $\phi_{\nu}$ we perform high resolution
numerical calculations using a 1-D (spherically symmetric) hydrodynamics code.
This code, written by us, borrows from the VH-1 code (see
http://wonka.physics.ncsu.edu/pub/VH-1/index.html) and uses the same piecewise
parabolic method (PPM) with a Lagrangian step followed by a remap onto the
fixed grid.  We include the non-equilibrium ionization of the gas and
radiative cooling appropriate to the ionization.  For these calculations the
ionization, recombination and radiative cooling rates have been generated
using the \citet[][plus updates]{RS77} codes.  Care has been taken to assure
that the remapping step conserves the mass in each ion stage. We have carried
out resolution studies and have found that the results converged at high
resolution.  The spectra used in the results that follow are for our highest
resolution (0.07 pc for the $n_a = 0.1\,$cm$^{-3}$ case, where $n_a$ is the
average ambient preshock hydrogen nucleus density of the medium into which
supernovae expand).  The evolution of a remnant has been followed until nearly
all the energy has been radiated and a small fraction of the explosion energy
remains in thermal energy.  The spectrum is then generated using the
temperature, density and ionization profiles as a function of remnant age.  We
have done some runs in which thermal conduction was included (using an
operator splitting technique) with no inhibition of conduction other than the
limitation imposed by saturation \citep{CM77}.  We have found the
differences in the resulting spectra to be relatively small and therefore do
not present results for these cases in this paper.  We have done runs for a
variety of ambient density cases, $n_a = 0.04$, 0.1, 0.3 and 1.0 cm$^{-3}$.
We assume an explosion energy of 10$^{51}\,$ergs in all cases.  We do not
include any magnetic field effects; this assumption is discussed below.

Having generated $\phi_{\nu}$ (and thus the mean emissivity, $\meps$, assuming
$S E_{\rm SN}$ is known) we then need a model for the opacity of the medium in
order to calculate the mean intensity, $J_{\nu}$.  The true opacity of the ISM
to ionizing photons is clearly extremely complicated.  It depends on the
distribution of neutral hydrogen column density on scales ranging from
$10^{18}\,$cm$^{-2}$ to $10^{20}\,$cm$^{-2}$, including the correlations among
clouds and local large scale structures.  To create a realistic model for the
opacity of the ISM is thus well beyond the scope of this paper.  Nevertheless,
any given cloud that is to be subject to the ionizing field generated by hot
gas will receive radiation from sight lines that will pass through a range of
cloud columns present in the medium.  For this reason we adopt a model for the
opacity that is as simple as possible while still allowing for an
interspersion of emitting and absorbing material, as must be the case in the
ISM.

Our approach is to assume a uniform medium in the sense that the clouds,
assumed to be of some typical optical depth, $\tcn$, are distributed randomly
in the hot gas.  The optical depth due to these clouds at frequency $\nu$ is  
\begin{equation}
\tau_{\nu} = \nlos\left(1-e^{-\tcn}\right)
\end{equation}	
\citep[][see the Appendix]{BF69}, where $\nlos$ is  the expected
number of clouds along the line of sight.  The average value of the mean
intensity in the disk and halo is
\begin{eqnarray}
\jsnb & = &\frac{1}{\ton}\int_0^{\ton}d\tau_\nu\;\frac{1}{4\pi}
	\int I_\nu d\Omega,\\
     & = &\frac{\epna}{4\pi\ton}(1-\eta_\nu)
\label{eq:jbar}
\end{eqnarray}
where $\ton$ is the optical depth through the full disk and halo, 
\beq
	\eta_\nu=\frac{1}{\ton}\left[\frac{1}{2}+E_3(\ton)\right],
\label{eq:eta}
\eeq
is the mean escape probability, and $E_3(\ton)$ is an exponential integral.
As shown in the Appendix, this value of the mean intensity is approximately
equal to the expected value of the mean intensity in the intercloud medium,
provided it has a substantial filling factor.

Thus our model for the mean intensity incident on a cloud depends on five
parameters: the cloud optical depth $\tcn$, the disk optical depth, $\ton$,
the SN explosion energy, $E_{\rm SN}$, the supernova rate per unit area $S_A$,
and $\phi_{\nu}$.  The optical depths are related to the column densities by
$\tcn = \NHoc\sigma_{\nu}$ and $\ton =\NHop/\NHoc [1 -
\exp(-\NHoc\sigma_\nu)]$ (eq \ref{eq:ton}), where $\NHoc$ is the HI column
through the standard cloud and $\NHop$ is the total HI column of the disk and
the halo perpendicular to the disk plane.  The values for the column densities
are discussed in \S 3 below.  For $S_A$ we use the results from \citet{MW97},
$S_A = 3.8\times10^{-5}\,$kpc\ee\ yr\e\ at the Solar Circle. However, for the
line of sight toward HD 93521 we find it necessary to effectively reduce this
number slightly in order to match the observed H$\alpha$ intensity. In
general, $E_{\rm SN}$ and $\phi_{\nu}$ are not independent, since the
explosion energy will affect the evolution (temperature, ionization, etc.) of
the remnant which in turn determines $\phi_{\nu}$. Nevertheless, the
separation of the two quantities emphasizes the fact that $\phi_{\nu}$ is the
efficiency with which the available energy is radiated over frequency.  Thus
our formulation could be extended to other potential sources of ionizing
radiation such as the conduction fronts surrounding evaporating clouds
\citep{MC77} or the interfaces in turbulent mixing layers \citep{SSB93}.

\subsection{The Model X-ray/EUV Spectrum}
\begin{figure}
\plotone{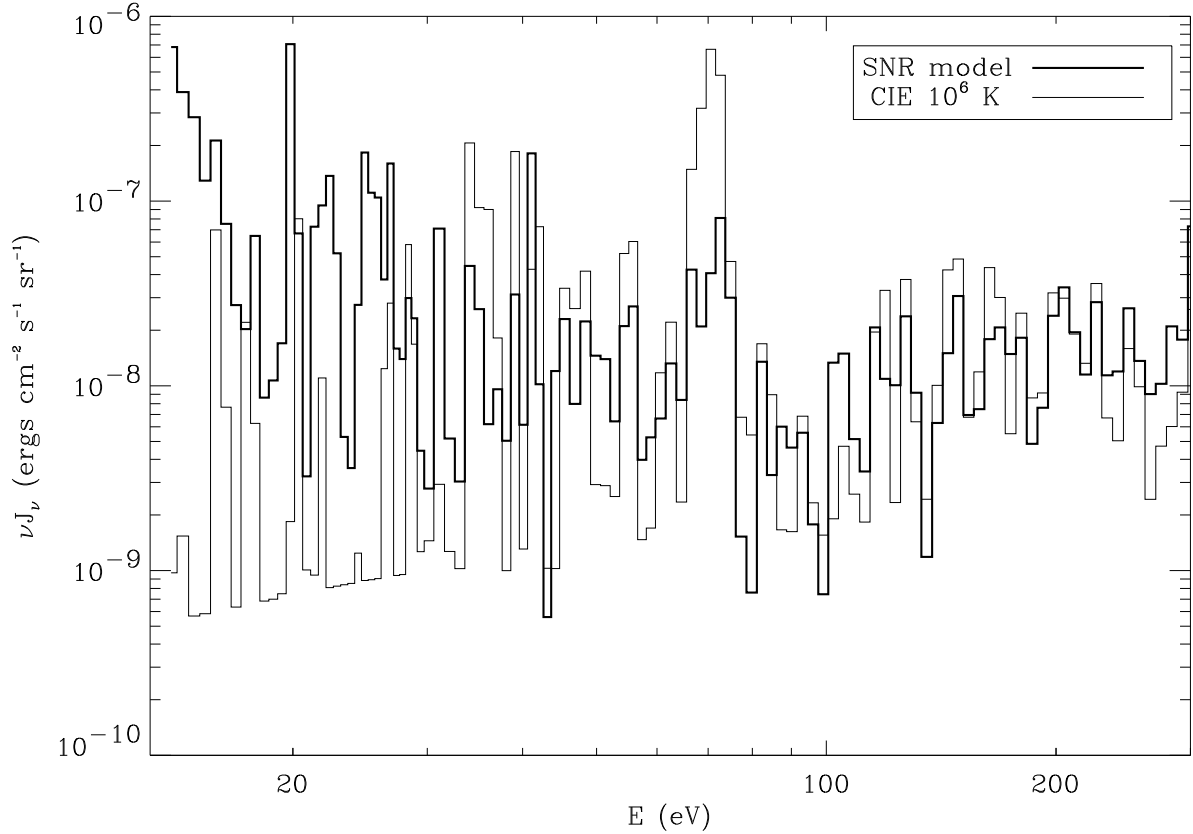}
\caption{Mean intensity incident on a typical cloud for
our model spectrum with ambient density 0.1 cm$^{-3}$ and no conductivity.
For comparison we also plot the mean intensity of a collisional ionization
equilibrium, $T = 10^6\,$K plasma with the emission chosen to match the
all-sky average for the Wisconsin B band observations. (The CIE spectrum does
not provide quite enough emission to match the all-sky average C band rate.)
\label{fig:spect}}
\end{figure}

Figure \ref{fig:spect} shows our model X-ray/EUV spectrum compared with a
collisional ionization equilibrium (CIE), $T = 10^6\,$K, unabsorbed spectrum.
The latter is scaled so as to match the all-sky average B band count rate
observed by the Wisconsin group \citep{MBSK83}. Of particular note are the
greater fluxes at low energies ($h\nu \sim 13.6 - 30$ eV) in the cooling model
as opposed to the CIE spectrum.  As can be seen from the spectrum, these photons
dominate the higher energy photons in producing ionization in interstellar 
clouds.

A great deal of effort has gone into modeling the observed diffuse soft X-ray
background (SXRB) \citep[e.g.][]{CA82,JK86,Sea98,SC98}. Our model has not been
created with the aim of matching the SXRB observations.  In particular, we do
not attempt to match the plane to pole variation in intensity or the variation
of the observed X-ray band count rate ratios with \NHI.  These characteristics
of the data apparently demand the existence of an irregularly shaped Local
Bubble and as such are inconsistent with any simple global model.  On the
contrary, we have calculated a time and space averaged emission spectrum that
should approximate the observed spectrum at high galactic latitudes, where the
lower \ion{H}{1} density allows us to see farther along a line of sight and
where there is clear evidence that the observed flux has substantial
contributions from distant emission \citep{Gea92}.

For comparison of our model results to the data, we need to specify all the
model parameters as described above.  Of particular importance is $\NHop$, as
it effectively determines the mean free path for the soft X-ray photons.  The
average disk column density at the solar circle is $6.2\times 10^{20}$
\sqc\ \citep{DL90}. However, the Local Bubble apparently has carved out a hole
in the \ion{H}{1} distribution, resulting in a substantially lower value of
$\NHop$ locally.  We use a value of $3.2\times10^{20}$ \sqc\ \citep{KF85} for
our comparisons at high latitude.  For the typical cloud column density, we
adopt  $N_{\Ho c} = 1.46\times 10^{19}$ \sqc, the mean value observed
for clouds along the line of sight towards HD~93521.  We use our standard
spectrum ($n_a = 0.1\,$\cc, no conduction, and $E_{\rm SN} = 10^{51}\,$ergs)
to determine $\phi_{\nu}$.  
In order to match the observed \Ha\ intensity
toward HD~93521, we find it necessary to reduce $S_A$ by a factor 0.68 (\S
\ref{sec:photo}).  In other words, the average supernova rate $S_A$ at the
Solar Circle creates enough cooling hot gas to produce somewhat more
hydrogen ionization than is observed toward the particular line of sight
to HD 93521; this clearly demonstrates that the ionizing power of supernova
remnants is significant at high latitude. A slight reduction in $S_A$ 
along this line of sight, or of the radiation field along this line of sight,
is quite reasonable in view of the
observed variations in the soft X-ray background \citep[e.g.][]{Sea97}.

The most important comparison we can make with the data is a spatially
averaged measurement at low energy.  For this purpose, the Wisconsin B and C
band data are the best choice.  From the publicly available maps we find that
for $|b| > 45$\arcdeg\ the average count rates for the B and C bands are 63
and 184 counts s\e, respectively. For the line of sight towards HD~93521, the
count rates are 84 and 210 counts s$^{-1}$.  There is a fair amount of
scatter, with the B band rate ranging from 18 to 134 and the C band rate from
62 to 302.  The C/B ratio shows somewhat less scatter, ranging from 1.7 to
8.2.  For the parameter choices detailed above we find a B band rate of 70 and
a C band rate of 202 counts s$^{-1}$.  Given that we have made no attempt to
vary the parameters of our model to fit the SXRB data, we find this agreement
remarkable.

\subsection{Soft X-ray Emission from External Galaxies}
\label{ssec:soft}

Another test of our model is to compare it with the soft X-ray emission from
external galaxies.  This comparison has the advantage that local variations
are averaged out, but the disadvantage that various conditions affecting the
strength and spectrum of the emission (e.g.\ $n_a$, $S_A$) could be
significantly different in other galaxies.  We estimate the X-ray luminosity
of the Galaxy using the average conditions at the solar circle; thus, we use
$\NHop = 6.2\times 10^{20}\,$\sqc\ for the disk thickness, $S_A=3.8\times
10^{-5}$ kpc\ee\ yr\e\ for the SN rate per unit area, and 530 kpc$^2$ for the
effective area of the disk \citep{MW97}.  Using equations (\ref{eq:jbar}) and
(\ref{eq:eta}), we find that the total luminosity in ionizing photons that
escape from the disk in this case is $1.1\times10^{40}$ ergs s$^{-1}$, while
the luminosity in X-ray photons ($E > 100$ eV) is $2.2\times10^{39}$ ergs
s$^{-1}$.  Note that our estimate for the ionizing luminosity is very
sensitive to our assumption that the emission and absorption are uniformly
mixed; our estimate for the X-ray luminosity also depends on this assumption,
although to a lesser extent (see the Appendix).

Observations of diffuse soft X-ray emission in external galaxies are quite
difficult.  Proper accounting for point sources, backgrounds and galactic
absorption are required \citep[hereafter CSM]{CSM96}.  The intensity or
emission measure estimates that result are not tightly constrained.  CSM
observed five galaxies (NGC 3184, NGC 4736, M101, NGC 4395 and NGC 5055) in
the R12 or R12L bands of ROSAT ($E \approx 100$-284 eV).  In all cases the
emission is sharply peaked near the center of the galaxy and drops off quickly
with galactocentric radius.  For our standard case ($n_a = 0.1\,$\cc, no
conduction) we find in the R12L band a rate of $550 \times10^{-6}$ counts
s$^{-1}$ arcmin$^{-2}$ (this is the standard unit for such ROSAT
observations).  Comparing this result to the surface brightness measured by
CSM at the boundary between rings 2 and 3 (which corresponds approximately to
the Solar Circle), we find that our result lies below the 95\% upper limits
for all the galaxies observed, and above the 95\% confidence lower limits for
all but two of the galaxies listed, namely NGC 4736 and NGC 5055.  The fact
that we get agreement with the observations, despite the large luminosity is a
result of the softness of our spectrum.  As has been pointed out previously
\citep{CM86}, the relative faintness of the observed soft X-rays from external
galaxies requires that much of the supernova energy is absorbed in the disk.
Our model provides a natural explanation of why that occurs. Most of the
radiated supernova energy is generated at low energies (see Figure
\ref{fig:spect}) and few of the photons escape.

\section{Photoionization Due to the Hot Gas Emission} 
\label{sec:photo}

Soft X-ray/EUV photons that are radiated by hot gas from SNRs will either be
absorbed or escape the galaxy.  Since the mean half-thickness of the disk is
$N_\Ho \sim 3\times10^{20}\,$cm$^{-2}$, nearly all of the ionizing photons less
energetic than 0.5 keV will be absorbed.  The surprisingly low soft X-ray
brightness observed for external galaxies (CSM) is an indication that a large
fraction of the X-rays generated by SNRs are being retained by the galaxies
and therefore should contribute to their ionization.  The photoionization rate
of species $i$ per unit area of disk is then
\beqa
\zeta_{iA}&=&\int_{{\nu_i}}^\infty d\nu(1-\eta_\nu)\left(\frac{\sigma_{i\nu}}
     {\sigma_\nu}\right)\frac{\epna}{h\nu},\\
       &\simeq&y_i\int_{{\nu_i}}^\infty d\nu \;\frac{\epna}{h\nu},
\eeqa
since $\eta_\nu\simeq 0$; here $\nu_i$ is the frequency threshold for
ionization of species $i$ and $y_i$ is the fraction of the photons absorbed by
species $i$.  For hydrogen, $y_\Hy$ is very close to unity since almost
every helium ionization results in the emission of a hydrogen ionizing photon
\citep{O89}; for helium, $y_\Heo\sim 0.6$ in weakly ionized regions
since hydrogen competes for the ionizing photons.  Defining
$\epias\equiv\int_{\nu_i}d\nu\epna/h\nu$, which is the rate of generation of
photons with $\nu\geq\nu_i$ per unit area, we can express the ionization
equation as simply
\beq
\zeta_{iA}=y_i\epias.
\eeq
Note that here and below we use the superscript (*) to denote units of photons
instead of units of energy; this should not be confused with the conventional
notation for absorption lines from excited fine structure states, such as
$\Cps$.  The ionization rate is thus almost independent of the morphology of
the \NHI\ distribution---it does not matter whether the \ho\ is in small,
optically thin clouds or large optically thick slabs (for helium, there can be
a weak dependence of $y_\Heo$ on the cloud morphology).  The ionization of
hydrogen determines the emission measure
\beq
\emppc=\frac{\ephas}{\alt_\Hy}=1.25\times 10^{-6}T_4^{0.8}\ephas
     ~~~{\rm cm^{-6}\ pc}
\label{eq:emppc}
\eeq
from equation (\ref{eq:ephas}).  The subscript ``pc'' on $\emppc$ 
indicates that it is measured in units of cm$^{-6}$ pc; $\ephas$ has units
photons cm$^{-2}$ s$^{-1}$; and $\emppc$ is the emission measure normal to the
plane of the galaxy and through the entire disk.  
Photoionization by emission from hot gas thus
sets a floor on the emission measure that should be observable on nearly all
lines of sight.  

	To relate the ionization rate to the properties of 
individual SNRs, we define
\beq
\phig\equiv\int_{\nu_i}^\infty d\nu\;\phi_\nu 
\eeq
as the fraction of the SN energy emitted above $\nu_i$ and
\beq
\frac{1}{h\nib}\equiv \frac{1}{\phig}\int_{\nu_i}^\infty d\nu\;\frac{
    \phi_\nu}{h\nu}
\eeq
is the inverse of the mean energy of ionizing photons for species $i$.  In
terms of these quantities, the number of photons emitted above $\nu_i$ by an
SNR is 
\beq
\calng(>\nu_i)=\frac{\phig E_{\rm SN}}{h\nib}=4.6\times 10^{61}\phig \left[\frac
{E_{\rm SN}}{10^{51}\ {\rm erg}}\right]\left[\frac {13.6\ {\rm 
eV}}{h\nib}\right] ,
\eeq
and the rate of generation of ionizing photons per
unit area is
\beqa
\epias&=&S_A\calng(>\nu_i)\\
    &=&1.26\times 10^6\left(\frac{S_A}{3.8\times
    10^{-5}~~{\rm kpc^{-2}\ yr^{-1}}}\right)
    \left[\frac{\calng(>\nu_i)}{10^{61}~~{\rm photons}}\right]
    ~~~~{\rm photons\ cm^{-2}\ s^{-1}}.
\eeqa
In Table \ref{tbl:phi} we illustrate the nature of the spectrum emitted by SNR
for the different model assumptions by listing the fraction of the SN energy
emitted in several different energy bands and the commonly used ionizing
photon ratio, $Q$(He$^0$)/$Q$(H$^0$).  This latter is the ratio of the total
number of He$^0$ ionizing photons to H$^0$ ionizing photons emitted.  The
energy bands are defined as follow: $\phi_\Hy$: 13.6--24.6 eV, $\phi_{\Heo}$:
24.6--54.4 eV, $\phi_{\Hep}$: 54.4--284 eV, $\phi_X$: $> 284$ eV.  The first two
are defined by the hydrogen and helium ionization edges.  The edge at 284 eV
is the carbon edge above which C and other elements are responsible for a
significant amount of the opacity.  The C edge also corresponds (not
coincidentally) to the cutoff in response in the Wisconsin C band and the
\emph{ROSAT} R1 and R2 bands.  It is clear that while there is a general trend
towards harder spectra for SNR expanding in higher ambient density, $n_a$, the
trend is not monotonic in all the energy bands.  This is due to fact that the
emission spectrum is line-dominated so that a few strong lines from ions of 
an abundant element can substantially affect the total emission within an
energy band.

Table \ref{SNRphot} lists the emission measures generated by our modeled SNRs,
together with characteristics of the ionizing spectrum, for several values of
the ambient density and thermal conductivity.  We have assumed $T=8000$~K in
evaluating the emission measure, and we have divided the value given in
equation (\ref{eq:emppc}) by 2 in order to give the value that would be
measured from the midplane of the disk.  Typically somewhat more that half the
energy of a SNR is emitted in ionizing photons, and the mean energy of these
photons is only about 20 eV.

The effects on the spectrum of varying conductivity are quite small, despite
the fact that the temperature at the center of the remnant is radically
different for the two cases.  When thermal conduction is included, the central
temperature flattens quickly and the thermal energy is shared with matter 
further out in the remnant.  The effects on the emission are small, however,
because the emissivity is sharply peaked towards the edge of the remnant where
the temperature and density profiles are little effected by conduction.

The effects of varying the ambient density are similarly small.  To first
order, the luminosity, spectrum, and $EM$ are independent of $n_a$. However,
as can be seen from Tables \ref{tbl:phi} and \ref{SNRphot}, there is a weak
trend in that higher ambient density leads to slightly harder emission
spectrum.  This is due to the fact that a supernova remnant evolving in a
higher density medium becomes radiative earlier, when the gas inside is at a
higher temperature.  Since the radiative phase is the period of greatest
luminosity for the remnant, this has an effect on the mean emissivity even
though the radiative phase lasts for only a small fraction of the remnant
lifetime.  The differences in the hardness of the spectrum for the different
cases can be seen both from the mean energy of ionizing photons and the
fraction of the emitted energy in ionizing photons.

It is noteworthy that while the ambient density in our models varies by a
factor of 25, the emission measure generated for the extreme cases differ by
only 34\%.  The intensity of an \Ha\ line in Rayleighs is related to the
emission measure by $I^*({\rm H}\alpha) = 0.445 \empc$~R for $T=8000$~K from
equation (\ref{eq:istar}). (Note that although the relation between $I^*$ and
$EM$ depends on temperature, the relation of $I^*$ to the underlying
emissivity $\ephas$ is almost independent of temperature---see eq.
\ref{eq:iperpstar}.) Our standard model gives $(\emppc/2)=1.2$~cm$^{-6}$~pc
for the half disk, which corresponds to a predicted intensity at $b=90\arcdeg$
of $I^*(\mha)=0.53$~R.  By comparison, \citet{R91} has suggested 1 R as a
``typical'' minimum value for the diffuse \Ha\ background at $b = 90\arcdeg$,
with the emission roughly following a $\csc b$ law.  Those conclusions were
based on a sampling of regions observed at moderately high latitudes using a
less sensitive instrument than the current instrument, WHAM (Wisconsin \Ha\
Mapper).  Even those data showed substantial variation in the value of $I \sin
b$, and more recent data reveal the existence of many regions of low emission,
well below the 1~R zenith value of the earlier estimate. One of these is the
direction towards HD~93521 discussed in detail below.  Thus, SNRs are capable
of accounting for the observed \Ha\ intensity in at least some directions in
the halo.  It is clear that the cooling hot gas in our model makes a
substantial contribution to the ionization of hydrogen in the warm ionized
medium of our galaxy and sets a lower limit on the ionization.

Another way of testing our model is to compare it to the ionization of other
elements.  Absorption line data provide particularly good tests since high
spectral resolution observations allow us to separate different velocity
components and thus model individual clouds.  A particularly useful example is
the line of sight towards HD 93521.

\section{The HD 93521 Line of Sight}
\label{sec:hd}
\subsection{The Data}
\label{sec:data}

The line of sight towards the halo star HD~93521 ($\ell = 183 \arcdeg, b =
62\arcdeg$), while in some ways quite unusual, reveals important information
about the background ionizing radiation field in the galaxy. The line of sight
is unusual in two ways. First, there appear to be no stars close enough to it
to be significant ionizing sources (with the exception, of course, of HD~93521
itself, which could ionize no more than one of the clouds, see below). Note
that it is not unusual for any given point in the galaxy to be far from O
stars, rather it is the fact that the entire 1.7 kpc line of sight is far from
any O stars that makes it somewhat exceptional.  This
can be seen by examining figure 16 (particularly 16a and 16b) of \citet{MC93}.
In that figure, they present the outlines of the extended \ion{H}{2} regions
resulting from the observed distribution of nearby O stars.  The line of sight
does not pass through any of the \ion{H}{2} regions, reflecting the fact that
no O stars are close to the line of sight.  [Note: in figure 16c the Sun and
the line of sight towards HD~93521 appear to be on the edge of an \ion{H}{2}
region. It is unclear which star might be responsible for this ionization
(perhaps $\zeta$ Oph) and it is unlikely to be capable of ionizing clouds
towards HD~93521.]  Secondly, observations by \citet[hereafter SF]{SF93} show
that there is extremely little cold neutral gas along the line of sight.  Thus
we have in this line of sight an excellent opportunity to observe ionization
of warm diffuse gas by non-stellar sources without confusion by interspersed
cold gas.

The absorption line data for HD~93521, including lines of \ion{S}{2},
\ion{S}{3}, \ion{C}{2}$^*$, and several other lines, reveal a number of
velocity resolved features, or ``clouds'', all indicating partial ionization
(SF).  Complementing the absorption line data are new observations using the
Wisconsin \Ha\ Mapper \citep[WHAM;][]{Rea98} of diffuse emission in \Ha,
[\ion{N}{2}]$\,\lambda$6584, and [\ion{S}{2}]$\,\lambda$6716 \citep{Pea99}.
The velocity components in the absorption lines split roughly into two groups
of 4 clouds each, the ``fast'' clouds with $-70\; {\rm km\ s}^{-1} < v_{\rm
LSR} < -30\; {\rm km\ s}^{-1}$ and the ``slow'' clouds with $-20\; {\rm km\
s}^{-1} < v_{\rm LSR} < 10\; {\rm km\ s}^{-1}$.  The emission lines cannot be
resolved into individual cloud components at this time, but they do split up
into broad ``fast'' and ``slow'' components that correspond to the absorption
line groups.  \citet{Hea99} find that for the slow clouds, $I^*({\rm H}\alpha)
= 0.22 \pm 0.06\,$R, and for fast clouds, $I^*({\rm H}\alpha) = 0.16 \pm
0.03\,$R.  \citet{Pea99} find $I^*$([\ion{N}{2}]$\,\lambda 6584)=0.22$~R for
the slow clouds and 0.29~R for the fast ones; they also find
$I^*$([\ion{S}{2}]$\,\lambda 6716)=0.24$~R for the slow clouds and 0.19~R for
the fast ones (see Table \ref{tbl:lines}).

\subsection{Modeling the Ionization}

In order to calculate the ionization of clouds subjected to the SNR-generated
flux of our model (Figure \ref{fig:spect}) we make use of the
photoionization/thermal equilibrium code CLOUDY \citep{fer96}. The flux has
been scaled by reducing $S_A$ by 0.68 in order to give the observed value for
the \Ha\ intensity.  In order to approximate the effect of isotropic, diffuse
radiation using CLOUDY, which is designed for point source calculations, we
use the ``illuminate'' command with the parameter value 60, which simulates
flux coming into the cloud face at an angle of 60 degrees.  We also use the
``case b'' command to avoid the artificial excitation of \Ha\ emission due to
absorption of FUV background photons at Lyman $\beta$.  We use a fixed
temperature and have explored a range from 6000 K to 9000 K.  Fixed $T$ avoids
the complications of heating/cooling balance that involve the amount of dust
heating (particularly due to PAHs) \citep{BT98}, which is only crudely
included in the code, and the uncertain details of the elemental abundances.
We discuss heating further below.

Although the usual approach is to assume that individual velocity features are
physical ``clouds'', we examined the possibility that the features were
actually formed from collections of smaller clouds or ``cloudlets''.  This
affects the opacity of the medium and thus the hardness of the radiation field
incident on a typical cloud, but it does not affect the emission measure along
the line of sight since nearly all ionizing photons are absorbed somewhere in
the disk.  The difference between many small cloudlets and a few big clouds is
greatest near the Lyman limit, where for small cloudlets the incident flux is
much less.  We have constructed two models to try to differentiate between
these two cases.  The ``cloudlet'' model has cloudlet column densities of
$N_\Ho = 7\times10^{17}\,$cm$^{-2}$.  The simple cloud model has $\NHo =
1.46\times 10^{19}\,$cm$^{-2}$, the mean for the observed clouds towards
HD~93521. The only data we have found that differentiates these two fairly
extreme cases from each other is the ratio of $N_\Spp/N_\Sp$.  This ratio has
unfortunately only been observed for the fast clouds, but if we assume that
the fast and slow clouds are at least geometrically similar (which might be
doubtful, see below), then our results favor larger clouds over cloudlets.
This is best seen by forming a ratio that is insensitive to the pressure of
the clouds but still sensitive to the cloud size,
$(N_\Spp/N_\Sp)\times(N_\Cps/N_\Su)$, where $N_\Su\simeq N_\Sp + N_\Spp$ to
good accuracy.  The lack of sensitivity to pressure derives from the fact that
$N_\Spp/N_\Sp$ goes as 1/$n_e$ (for a fixed input spectrum), while
$N_\Cps/N_\Su$ goes as $n_e$.  Comparing this ratio for both cloud size
assumptions, we find that the cloudlet model results in too low a ratio, $\sim
0.0007$, whereas the larger clouds yield a ratio of $\sim 0.002$, close to the
observed values.  For this reason, and because of the various conceptual
difficulties inherent in the cloudlet picture, we have focussed our attention
on the more standard cloud picture (i.e.\ velocity features corresponding to
spatially coherent clouds) in our comparisons with the data.  It should be
noted, however, that lacking the SIII line observations for the slow clouds,
we are unable to differentiate between the cloud size models for those clouds.

Up to this point we have not discussed the ionization due to HD~93521 itself.
According to \citet{VGS96}, HD~93521, an O9.5V star, should have a luminosity
in ionizing photons of $2.40\times10^{48}\,$s$^{-1}$.  This then implies that
the star can create enough ionization to produce $I^*(\mha) = 
9320/r_{\mathrm{pc}}^2$ Rayleigh from equation (\ref{eq:iperpstar}), where
$r_{\mathrm{pc}}$ is the distance of the cloud from the star in pc.  We do not
have any direct information on the distance between HD~93521 and the cloud
nearest to it.  HD~93521 is determined to be at a height above the plane of $z
\simeq 1500\,$pc, well above scale height of neutral and even ionized gas.
From this we estimate that the contribution of HD~93521 to the \Ha\  observed
towards it is $\lesssim 10$\%.  In any case, since all the clouds observed by
SF are optically thick in the EUV, the first cloud along the line of sight
will absorb all the ionizing radiation and the other clouds are not affected
by HD~93521.  We do not include its contribution to the ionization in our
calculations.

\subsection{Model Results \label{sect:modres}}
Our model results for both the absorption lines and emission lines, combined
into ``fast'' and ``slow'' cloud groups are compared with the observations in
Table \ref{tbl:lines}.  We use results from a series of CLOUDY runs with the
same input spectrum (SNR model with $n_a = 0.1$, $E_0 = 10^{51}$ ergs, no
conduction) and pressures ranging from $P/k = 1000$--$2\times10^4$\ccK.  The
model value for the observed column densities and intensities for each cloud
is determined by finding which pressure value led to the best match with the
observed \CIIe/\ion{S}{2} line ratio at the observed value of $N_\Sp$.  We
then use the results of the best match pressure run for determining the values
for the other observables.  We choose to peg our results to the
\CIIe/\ion{S}{2} line ratio because it is one of the best determined
observationally for the line of sight.

Because we fix the other quantities calculated in our models by their values
at the observed $N_\Sp$ values for each cloud, the model and observed values
for $N_\Sp$ are identical.   We also match the total line of sight (fast+slow)
value for \Ha\ intensity by scaling the input spectrum as described in
\S\ref{sec:photo}.  The model values for \NCIIe\ are also very close to the
observed values, as a result of our fixing the pressure value by finding the
best match for the \CIIe/\ion{S}{2} line ratio as described above. Thus the
first three lines of the table consititute inputs for the model.  For the rest
of the table, the model results are truly outputs, not having been fixed to
match the data.  For these results (as well as for $N_\Cps$ for completeness)
we have presented ranges for the model results corresponding to a (fixed)
cloud temperature of 6000 K and 8000 K respectively.  This range of values
might correspond to the temperatures in the neutral and ionized portions of
the clouds as we discuss below.

One of the most confusing aspects of the absorption line data is the
apparently large neutral column densities of the clouds, as seen in either
\ion{H}{1} or \ion{S}{2}, coupled with large ionized columns, as inferred from
the relatively high columns of $\Cps$, which is mainly excited by collisions
with electrons in diffuse, $n\approx 0.2$~cm\eee\ gas.  This is especially
true for the slow clouds. If one interprets each velocity component as being
due to a partially ionized cloud, these data require either a very large
ionizing flux and/or a high thermal pressure.  As we discuss in more detail
below, this is because $N_\Cps/N_\Ho \sim x n/(1 - x)$, where $x\equiv
n_\Hp/n$ is the fractional ionization of hydrogen.  The observed values of
$N_\Cps/N_\Ho$ imply high values for $x$ and/or $n$ for the clouds at the same
time that $N_\Ho$ is large ($3\times 10^{18}- 2\times 10^{19}\,$\sqc).

\begin{figure}
\plotone{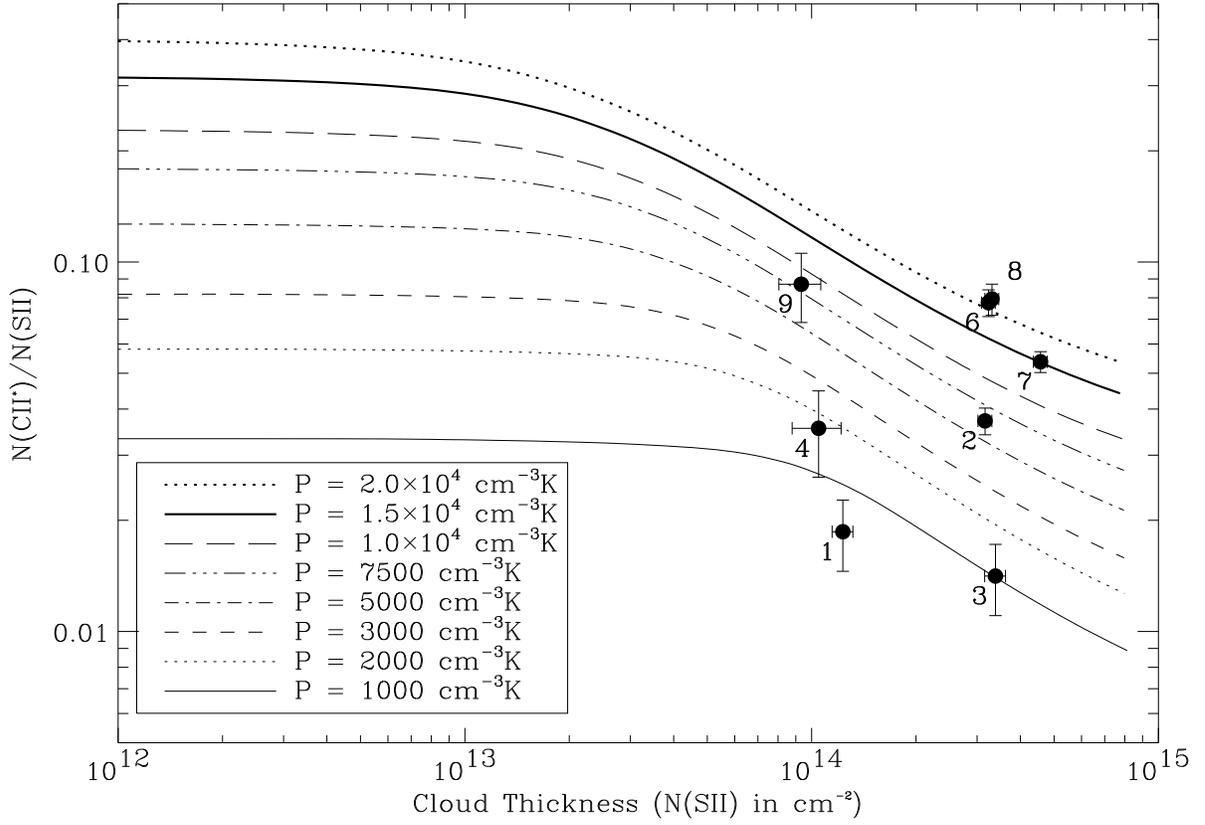}
\caption{Ionization created by our model spectrum incident
upon clouds at different thermal pressures vs.\ thickness of the cloud.  It is
clear that slow clouds in particular (nos.\ 6--9) demand high pressures to be
consistent with both the \Ha\ and absorption line observations.
\label{fig:CIIeSII}}
\end{figure}

The \Ha\ data taken with WHAM \citep{Hea99}, however, provide tight
limits on the ionizing flux.  For an assumed temperature of 8000 K, their data
imply an emission measure of 0.85 cm$^{-6}\,$pc for this line of sight.  With
the added information from the emission measure ruling out a very large
photoionizing flux, we are faced with the prospect of large thermal pressures
in the slow clouds.  This is demonstrated in figure \ref{fig:CIIeSII}, where
we plot the $N_\Cps/N_\Sp$ ratio based on calculations using our model flux.
The flux, which is assumed to be the same for all the clouds, has been
adjusted to match the total \Ha\ intensity observed for the line of sight.
For our standard model this required reducing the flux by a factor of 0.68.
The $x$-axis is the total thickness of a cloud, as opposed to depth into a
cloud.  The points are labelled as in SF but with cloud no.\ 5 (the smallest
and most poorly detected component) excluded.  The fast clouds are numbered
1--4 while the slow clouds are 6--9.  We are able to match the observed ratios
fairly well, although rather high thermal pressures are demanded for the slow
clouds, $P/k \approx 7500-2\times10^4\,$\ccK.  There is evidence that in
regions as diverse as the fairly quiescent hot gas of the Local Bubble and the
cold neutral gas at greater distances \citep{JJL83}, high thermal pressures,
$P/k \sim 10^4-2\times10^4\,$\ccK, do exist in the ISM.  A difficulty with the
high value of the inferred pressure, however, is that it could lead to more
cooling than can be balanced by the known available heating sources at $T\sim
7000$ K, which would transform the cloud from the warm phase to the cold
phase.  We discuss the problem with the heating/cooling balance further below
in \S 6.2.

The reason for the high pressures demanded for the slow clouds can be seen
analytically \cite[see also][for a more extensive discussion]{MS99}.
Under the assumption of uniform ionization and temperature (as used by SF), the
ratio of \Ha\ intensity to $\Cps$ column density gives the hydrogen ionization
fraction, $x$,
\begin{equation}
\frac{I^*(\mha)}{N_\Cps} = \frac{\alpha_{\mha}(T) A_{21}}
{10^6\,\acg \gamma_{12}}\; x,
\label{HaNCIIe}
\end{equation}
where $\alpha_{\mha}$ is the effective recombination coefficient for emission
of \Ha\ (see the Appendix), $\acg$ is the abundance of gaseous carbon,
$\gamma_{12}$ is rate coefficient for excitation of C$^+$ to the $J=3/2$ level
and $A_{21}$ is the downward transition probability.  $I^*(\mha)$ is assumed
to be in Rayleighs and gaseous carbon is assumed to be entirely singly
ionized. Applying equation (\ref{HaNCIIe}) to the data towards HD~93521 (and
assuming a temperature of 8000 K) yields $x = 0.37\pm0.075$ and $x =
0.14\pm0.038$ for the fast and slow clouds, respectively.  (Note that the
temperature of 8000 K was used here rather than the 6000 K of SF because of
results from recent emission line measurements that we discuss below.)  This
estimate of the ionization can then be used with the observations of the
\ion{H}{1} column densities to get constraints on the pressure.  Assuming that
H and He are equally ionized we derive
\begin{equation}
\frac{N_\Cps}{N_\Ho} \approx 
\left(\frac{P}{kT}\right)\frac{\gamma_{12}}
{A_{21}} \acg \left(\frac{x}{1 - x^2}\right).
\label{pcon}
\end{equation}
Using our results from above this implies (again assuming $T = 8000\,$K) $P/k
= 1120 \pm 340$ and $9220 \pm 2770$ for the fast and slow clouds,
respectively, where most of the uncertainty derives from the $\sim20$\%
uncertainty in the \Ha\ intensities. For the more realistic case in which
ionization varies with depth into the cloud, as in our model, the value of $x$
derived in equation (\ref{HaNCIIe}) is weighted by $n_e$, thereby emphasizing
the most highly ionized part of the cloud.  As a result, the true mean
ionization will be less than that estimated assuming uniform ionization.
Equation (\ref{pcon}) carries with it a different weighting of $x$,
introducing more uncertainty in the determination of $P$.  Again the effect is
to emphasize regions of higher ionization resulting in an \emph{underestimate}
of the pressure.  Despite these limitations we infer that the fast and slow
clouds are markedly different in their mean ionization level and their
pressures.  The slow clouds have high thermal pressures, $\sim10^4$ \ccK . 
The fast clouds are more highly ionized and have lower thermal pressures,
$P/k \sim 2000$ \ccK , more typical of diffuse interstellar clouds
\citep{JJL83}.  

As detailed above, we have no difficulty matching the observed \Ha\
intensities towards HD~93521.  The other emission line data, observations of
[\ion{S}{2}] $\lambda$6716 and [\ion{N}{2}] $\lambda$6584 \citep{Pea99}, are
more problematic, however.  These data are consistent with the trend recently
discovered in the new WHAM data \citep[e.g.][]{Hea99} of both
[\ion{S}{2}]/\Ha\ and [\ion{N}{2}]/\Ha\ to increase to high values as \Ha\
decreases and as galactic latitude increases.  The values for these ratios are
a factor of 3 or more larger than is typical for diffuse emission in the
galactic plane.  The [\ion{N}{2}]/\Ha\ ratio is a good temperature diagnostic
since the ionization potentials of hydrogen and nitrogen are within 1 eV of
each other.  The reported values for $I(6584)/I$(\Ha) of  1 and 1.8 (for the
slow and fast clouds respectively) demand high temperatures, $T \approx 8000$
K and $T \gtrsim 9000\,$K.  Another temperature diagnostic is the
$N_\Cps/I$([\ion{S}{2}]) ratio, because $N_\Cps$ is relatively insensitive to
$T$ whereas [\ion{S}{2}] increases rapidly with rising $T$.  Since both C$^+$
and S$^+$ are expected to be the dominant stages of ionization in WIM/WNM gas,
both quantities in the ratio go as $\sim n_e n_\Hy$ and the density dependence
cancels in the ratio.  In contrast to [\ion{N}{2}]/\Ha, however, the observed
values for $N_\Cps/I$([\ion{S}{2}]) indicate relatively low temperatures, $T <
6000$ K for the slow clouds and $T \approx 7000$ K for the fast clouds.  A
potential solution to this seeming contradiction is to have substantially
lower temperatures in the neutral regions of the clouds relative to the
ionized regions. $I(6584)$ and $I$(\Ha) are weighted towards the ionized part
of the cloud (both the H$^+$ and N$^+$ densities are highest there), while
$N_\Cps$ and $I$([\ion{S}{2}]) have greater contributions from the neutral
part of the clouds.  As we noted above, we are unable to model the thermal
structure of the WIM/WNM accurately using CLOUDY because it does not yet
include PAH heating accurately.  Calculations by \citet{Wea95a} indicate that
temperatures in the WNM range from $\sim 5000$--9000 K depending on various
model parameters including the assumed pressure.  Higher pressures lead to
lower temperatures which is consistent with the observation that the high
pressure, slow clouds have a high $N_\Cps/I$([\ion{S}{2}]) ratio while the
lower pressure, fast clouds have a lower $N_\Cps/I$([\ion{S}{2}]) ratio.  In
summary, it appears that the slow clouds have $T\sim 5000$ K in the more
neutral regions and $T\sim 8000$ K in the more highly ionized regions, while
the fast clouds are characterized by $T\sim 7000$ K in the more neutral
regions to $T\sim 9000$ K in the more ionized regions.

One additional output from our modeling of the clouds is the filling factor of
cloud material.  This is more clearly stated in terms of the total line of
sight distance occupied by the clouds as determined by their densities and
column densities.  For our ``best fit'' models as described above we find that
the clouds account for only 100--150 pc of the $\sim 1700$ pc line of sight
towards HD~93521.  Thus only 6--9\% of the line of sight is occupied by the
warm ionized and neutral gas.  It is likely that this is not evenly
distributed but rather that there are more clouds near the plane.
Nevertheless, even assuming the clouds are confined within 1 kpc of the
galactic plane their filling factor would only be 9--13\%.  Thus for
this low column density, high latitude line of sight, warm diffuse gas 
appears to have a low filling factor.

\section{Helium Ionization and the X-ray Opacity of the Halo}

A possible constraint on the nature of the ionizing flux in the halo comes
from the recent study of the halo's X-ray opacity  by \citet[hereafter
A\&B]{AB99}.  A\&B inferred the opacity $\tau$ in Galactic gas needed to
account for the X-ray spectra of a number of galaxy clusters (including 13 at
high latitude) observed with \emph{ROSAT}.  In each case, they determined an
effective $\Ho$ absorption column density $\NHX$ based on the assumptions that
the absorbing gas is neutral and contains 10\% helium by number,
\beq
\tau\equiv \NHX(\sigma_\Ho+0.1\sigma_\Heo),
\eeq
where the cross sections, $\sigma_{\Ho}$ and $\sigma_{\Heo}$ are evaluated at
the energy appropriate to the X-ray observations, 0.25~keV.  They then
compared these column densities with those determined directly from 21 cm
observations, $N_\Ho$ (which they termed $N_{\mathrm{H, 21 cm}}$).  Because
the values of $\NHX$ make no allowance for ionized gas, which presumably
contains some neutral and once ionized helium, one would expect $\NHX$ to be
substantially larger than the actual $N_\Ho$.  What they found instead was
that for the low column, high latitude directions, $\langle \NHX/N_\Ho\rangle
= 0.972 \pm 0.022$. 

As A\&B pointed out, this result is difficult to understand.  The two column
densities are related by
\begin{equation}
\NHX(\sigma_\Ho+0.1\sigma_\Heo)=N_\Ho\sigma_\Ho+N_\Heo\sigma_\Heo
	+N_\Hep\sigma_\Hep,
\label{eq:nhx}
\end{equation}
where at the typical energy of the observations (0.25 keV), the cross sections
are $\sigma_{\Ho} = 1.08\times10^{-21}$ \sqc, $\sigma_{\Heo} =
2.81\times10^{-20}$ \sqc, and $\sigma_{\Hep} = 2.04\times10^{-20}$ \sqc.
Following A\&B, we define $C\equiv \NHX/N_\Ho$.  Let $x_\He$ be the helium
abundance relative to hydrogen, and let $\chi_\Hp\equiv N_\Hp/N_H$, $\chi_\Hep
\equiv N_\Hep/N_\He$, etc, be the fractional ionizations averaged along the
line of sight.  Inserting the values of the cross sections, we obtain
\beq
C=0.278+\frac{0.722}{1-\chi_\Hp}\left(\frac{x_\He}{0.1}\right)
	\big(1-0.274\chi_\Hep-\chi_\Hepp\big).
\label{eq:C}
\eeq
A\&B adopted $\chi_\Hp=0.27$ and $\chi_\Hep=0.5\chi_\Hp$ for the ISM at the
solar circle, which gives $C=1.23$ for a helium abundance $x_\He=0.1$.  If the
helium abundance were $x_\He=0.09$, as suggested by \citet{Bea91} in their
analysis of the Orion Nebula, then $C=1.14$.  Neither value comes close to
matching the observed value, $C=0.972$.

Our analysis of the line of sight toward HD 93521 gives a significantly lower
hydrogen ionization than adopted by A\&B, $\chi_\Hp=0.18$, but a comparable
helium ionization, $\chi_\Hep=0.15$.  The column density of doubly ionized
helium is negligible, $\chi_\Hepp = 0.0017$.  The difference in
the hydrogen ionization fractions is due in part to the fact that the fully
ionized HIM does not enter into our analysis, whereas it does contribute to
pulsar dispersion measures \citep{Wea95b} and thus to the value of $N_\Hp$
used by A\&B, as they recognized.  Our lower hydrogen ionization gives lower
values for $C$, 1.12 and 1.04 for $x_\He=0.1$ and 0.09, respectively. In view
of the many possible systematic effects that could enter into determining $C$
observationally, we do not regard the discrepancy between our model and A\&B's
observation as significant. 

\section{Discussion}
\subsection{The Calculated Spectrum}
The main results of this paper involve the ionization caused by cooling hot
gas from supernova remnants.  Thus questions of the accuracy of the spectra we
have produced are of central importance.  There are two concerns in this
regard: the supernova remnant evolution model and the plasma emission model.

It is clear that our SNR models are highly simplified.  In reality, SNRs
evolve in a very inhomogenous medium that contains material with a wide range
of densities, ionization states, magnetic field strengths and temperatures.
In the case of type II and type Ibc supernovae, the stellar progenitor will
have shaped the medium into which the remnant evolves.  The magnetic field in
the ISM can have important influences on the radiative properties of remnants
by limiting the compression in the cold shell that is formed during the
radiative phase \citep{SC92,SC93}.  Supernovae in OB associations combine to
form superbubbles that can grow to kpc size, sometimes venting their hot gas
into the halo.  We justify the simplifications we have made by appealing to
our results.  Despite the drastically different temperature and density
profiles that we see for the range of parameter values we explore ($n_a =
0.04$--1.0 \cc; conduction turned on or off), the time-averaged spectra are
not radically different.  Over its lifetime, the hot gas radiates its energy
primarily in the same lines in all cases, though differences in ionization and
temperature shift the balance from one set of lines to another for the
different cases.  Because we are concerned with the time and volume averaged
spectrum, the substantial differences between remnants at any given time in
their evolution is smoothed out.  Thus, while a strong magnetic field can
cause a delay in the radiation from a SNR, we do not expect the time and space
averaged spectrum to be substantially different from the spectrum for the no
field case.

The reliability of the plasma emission code that we use, the ``Raymond \&
Smith'' code, is currently an area of active study.  Soon more up-to-date and
detailed plasma emission codes will become available \cite[see][]{B99}. The
computed emission spectra and even the total cooling rates of hot plasmas
using these codes may differ significantly from those calculated using the
R\&S code.   However, for the reasons discussed in the preceding paragraph,
the photoionization rates of the time and space averaged spectra will likely
not be substantially changed. 

A potentially more important limitation of the calculations we have presented
is the lack of inclusion of interface radiation generated in the boundaries
between cold and hot gas.  Both thermal evaporation \citep{MC77} and turbulent
mixing \citep{SSB93} at those boundaries could lead significant enhancement of
cooling in those regions as well as an emission spectrum that is spectrally
different from those that we have presented here.  It is unclear at this time
what fraction of the energy radiated by SNRs could be emitted in such
interface regions.  This is a worthy subject for future investigations.

\subsection{Ionization and Thermal Balance in the Clouds}
An important difficulty for our modeling of the clouds towards HD~93521
concerns the heating/cooling balance.  As we mentioned in our description of
our model, we chose to use a constant temperature in the clouds to avoid the
sensitivity to gas phase elemental abundances and dust heating rates in our
cloud modeling.  Nevertheless, we have used CLOUDY, to examine the thermal
balance in the gas for some cases of interest.  We have found that, while at
low pressure ($P/k = 1000$--2000 \ccK) CLOUDY can produce temperatures in the
5000--8000 K range in the  ionized portion of the clouds, it has insufficient
heating to maintain temperatures above 1000 K in the neutral portions of the
cloud.   For the low pressure (fast) clouds, we attribute this to the grain
photoelectric heating model in CLOUDY.  Using the best grain heating rates
available, \citet{Wea95a} have shown that WNM (i.e, warm, $\sim 7000$ K,
neutral medium) can be maintained by dust and PAH heating at these pressures.
However,  the slow clouds  have large neutral hydrogen column densities, and,
from our modeling of the \CIIe, large pressures. Wolfire et al.\ also showed
that the WNM cannot be maintained at pressures above 10$^4$ \ccK,  as seems to
be demanded for some of the clouds.  This difficulty can be ameliorated, but
not entirely solved, if the slow clouds are broken up into small cloudlets.
The cloudlet model has the advantage that, because of the low column density
in an individual cloudlet, the ionization is more uniform.  As discussed  in
\S\ref{sect:modres}, more uniform ionization decreases the pressure required
to produce the observed $N_\Cps/N_\Ho$. However, Eq. (17) deomstrates that
even under the extreme assumption of uniform ionization, fairly high pressures
are demanded for the slow clouds, $P/k \sim 10^4\,$\ccK. Some additional
heating source seems to be needed to maintain these clouds at their observed
temperatures of $\sim 6000$--8000 K.  \citet{RHT99} have also presented
evidence from the WHAM survey for an additional heating mechanism in the halo.
Some possibilities for such a heating source include photoelectric heating by
small dust grains \citep{RC92}, magnetic reconnection heating
\citep{Ra92,BLN98} and dissipation of turbulence \citep{MB98}.

Another proposed source of ionization for the WIM is radiation from decaying
neutrinos \citep{Sc90}.  This source was seen to have the advantage that,
because neutrinos are essentially uniformly distributed throughout the Galaxy,
the ionization rate per unit volume is nearly constant.  Thus, clouds with
large \HI\ columns could be evenly and partially ionized throughout.  There
are several difficulties with this, however, that make neutrinos a very
unlikely candidate for the ionization of the clouds towards HD~93521.
First, we have identified a source of ionization associated with the {\it
observed} X-ray background that can account for essentially all the ionization
observed along this line of sight.  While the details in our model can be
improved upon by using more realistic models for the evolution of supernova
remnants and for the structure of the ISM, it is difficult to escape the
conclusion that the Galactic X-ray background has a low energy tail that can
produce significant photoionization.  Second, the photon energy from the
decaying neutrinos is estimated to be only about 13.7 eV \citep{Sc95}, which
is inadequate to account for the observed N$^+$ \citep{Pea99} or $\Spp$
\citep{SF93}.  Thus another source capable of creating these ions is needed in
any case, and this source will ionize hydrogen as well.  Finally, we note that
we find, in agreement with \citet{R91} and \citet{SF93}, that most of the
volume along the line of sight to HD 93521 is apparently empty.  In Sciama's
(1997) model, however, much of this volume is filled with gas that is
maintained in a fully ionized state by the decaying neutrinos.  This gas
should have detectable absorption lines, but none were found by Spitzer \&
Fitzpatrick.

\section{Conclusions}
Ionization in the warm diffuse interstellar medium of our galaxy is
substantially influenced by the soft X-ray/EUV emission from cooling hot gas
in supernova remnants.  In the galactic halo, hot gas emission is especially
important, dominating the ionization in some regions.  In this paper we have
presented calculations of the spectrum from the cooling hot gas and the
ionization that results.  We have found that: 

\begin{itemize}
\item the hot gas is capable of producing roughly 50\% of the EM observed for
a typical line of sight through the galactic disk, and more than enough to
account for that observed toward HD 93521;
\item our calculated spectrum is consistent with the observed soft X-ray
diffuse background at high latitudes;
\item the flux we calculate is also consistent with the soft X-ray emission
observed from other spiral galaxies;
\item we predict a low strength of the \ion{He}{1} $\lambda$5876 recombination
line along the HD 93521 sight line, consistent with low values seen in large
beam observations;
\item the ionization in our models allows us to match the $\Cps/\Sp$ and
the  $\Spp/\Sp$ ratios seen with absorption measurements 
along the HD~93521 line of
sight, though for the ``slow'' clouds large thermal pressures
($10^4-2\times10^4\,$\ccK) are required;
\item by varying the assumed temperature in our clouds from $\sim 6000-9000$ K,
we are able to match the observed emission line strengths of [\ion{S}{2}]
and [\ion{N}{2}], and predict the intensity of [\ion{O}{1}] $\lambda$6300;
\item with allowance for the uncertainty in the helium abundance and possible
systematic effects in the analysis of the data, we regard our value for the
X-ray opacity toward HD 93521 as being consistent with the value inferred by
\cite{AB99} for the halo;
\item by accounting for the observed ionization toward HD 93521 with a
theoretical extrapolation of a {\it known} source of ionization, the Galactic
X-ray background, we obviate the need to invoke exotic ionization mechanisms
such as decaying neutrinos \citep{Sc90}.
\end{itemize}

\acknowledgments
We are thankful to Ron Reynolds, Steve Tufte, and Nancy Hausen for sharing
their observational results with us prior to publication. We also thank Gary
Ferland for discussions regarding CLOUDY and sulphur dielectronic
recombination rates.  We wish to recognize the contribution of Xander Tielens
who participated in our early discussions of this project.  CFM gratefully
acknowledges support by an NSF grant (AST95-30480), a grant from the
Guggenheim Foundation, and a grant from the Sloan Foundation to the Institute
for Advanced Study. The authors acknowledge support from the NASA Astrophsical
Theory Program.

\appendix
\section{TRANSFER OF IONIZING RADIATION IN THE HALO}

In the absence of scattering, the equation of transfer is
\beq
\mu\frac{dI_\nu}{d\tau_\nu}=S_\nu-I_\nu,
\eeq
in standard notation; recall that $S_\nu=j_\nu/\kappa_\nu$ is
the source function.  We model the disk and halo of a galaxy
as having an emissivity $j_\nu$ and opacity $\kappa_\nu$
that depend only on the distance from the midplane, $z$.
Solving this equation for the flux, $F_\nu\equiv\int \mu I_\nu d\Omega$,
we find that the flux measured above the halo is
\beq
\fnp =  \fnm+4\pi\int_0^\ton S_\nu d\tau_\nu -4\pi\int_0^\ton J_\nu d\tau_\nu,
\label{eq:fnp}
\eeq
where $\fnm$ is the flux measured below the halo, $J_\nu=(1/4\pi)\int I_\nu
d\Omega$ is the mean intensity, and $\ton$ is the total optical depth of the
galaxy (measured normal to the plane).  Note that $4\pi\int S_\nu d\tau_\nu=
\int 4\pi j_\nu dz$ is the total emission per unit area.  The expected value
of $j_\nu$ is $\langle\epn\rangle/4\pi$, where the expected volume emissivity
$\langle\epn\rangle$ is defined in equation (\ref{eq:meps}).  We define the
emissivity per unit area as
\beq
\epna\equiv\int_{-\infty}^\infty dz\; \langle \epn\rangle
     =4\pi \int_0^\ton S_\nu d\tau_\nu.
\label{eq:epna}
\eeq
The mean escape probability---i.e., the fraction of the radiation
that escapes the galaxy---is
\beq
\eta_\nu=\frac{\fnp-\fnm}{\displaystyle 4\pi\int_0^\ton S_\nu d\tau_\nu}
      =\frac{\fnp-\fnm}{\epna}.
\label{eq:eta2}
\eeq
In terms of the mean escape probability, the mean value of the average
intensity is
\beq
\jnb\equiv\frac{1}{\ton}\int_0^\ton J_\nu d\tau_\nu
	=\frac{(1-\eta_\nu)}{\ton}\int_0^\ton  S_\nu d\tau_\nu
\eeq
from equation (\ref{eq:fnp}), so that
\beq
\jnb=\frac{(1-\eta_\nu)}{\ton}\left(\frac{\epna}{4\pi}\right).
\eeq

In order to calculate the mean escape probabability, we must make an
assumption about the spatial distribution of the emissivity and absorption.
We make the simplest assumption possible: The source function is constant
($j_\nu\propto\kappa_\nu$).  In that case, the flux escaping from the top of
the halo is
\beq
F_{\nu +}=\int d\Omega\int_0^\ton S_\nu e^{{-\tau_\nu/\mu}} d\tau_\nu
     =2\pi S_\nu\int_0^1d\mu\int_0^\ton e^{{-\tau_\nu/\mu}}d\tau_\nu.
\eeq
The magnitude of the flux escaping from the bottom of the halo is the same.
Evaluating the integrals and using equation (\ref{eq:eta}), we find
\beq
\eta_\nu=\frac{1}{\ton}\left[\frac 12 -E_3(\ton)\right],
\eeq
where $E_3$ is an exponential integral.  Although we have numerically
evaluated $E_3(\ton)$ in applying this equation in the text, it is sometimes
useful to have an approximate analytic expression.  The expression
\beq
\eta_\nu\simeq \frac{1}{1+2\ton}
\eeq
is accurate to within 20\%, whereas the corresponding approximation
for the absorbed fraction,
\beq
1-\eta_\nu\simeq \frac{2\ton}{1+2\ton},
\eeq
has the same accuracy provided $\ton>0.03$.  For ionizing photons with
energies somewhat above 1 Ryd, the escape probability is small; as a result,
the absorbed fraction is close to unity even if our assumption of a constant
source function is badly violated, whereas the escaping fraction is quite
sensitive to this assumption.  For X-rays, the optical depth of the disk and
halo is of order unity, so that the predicted X-ray luminosity of a galaxy is
somewhat sensitive to the assumption of a constant source function.

\subsection{Intensity of Recombination Lines}

Essentially all the ionizing photons that are absorbed in the disk and halo
ionize hydrogen; even those that initially ionize helium lead to the
production of a hydrogen ionizing photon \citep{O89}.  If we ignore ionization
by secondary electrons, the ionization equilibrium of hydrogen is governed by
\beq
\int_{\nu_\Hy}^\infty d\nu\;(1-\eta_\nu)\frac{\epna}{h\nu}
        =\alt\int_{-\infty}^\infty dz\; n_e n_p
	=\alt EM_\perp,
\label{eq:emperp}
\eeq
where $h\nu_\Hy$ is the ionization potential of hydrogen, $\alt\simeq
2.59\times 10^{-13}T_4^{-0.8}$ cm$^3$ s\e\ is the rcombination coefficient to
the excited states of hydrogen, and $EM_\perp$ is the emission measure normal
to the plane of the galaxy.  Ionization by secondary electrons becomes
dominant at large column densities \citep{Mea96}, but for the relatively small
column densities encountered in our model ($N\la 1.5\times 10^{19}$---see \S
\ref{sec:hd}), they contribute less than 2\% to the emission measure based on
calculations with CLOUDY.  In fact, the escape probability is small for almost
all the ionizing photons ($\eta_\nu\ll 1$), so the ionization balance can be
approximately described by
\beq
\ephas=\alt EM_\perp=7.99\times 10^5T_4^{-0.8}\emppc~~~
   {\rm photons\ cm^{-2}\ s^{-1}},
\label{eq:ephas}
\eeq
where
\beq
\ephas\equiv\int_{\nu_\Hy}^\infty d\nu\; \frac{\epna}{h\nu}
\eeq
is the ionizing photon luminosity per unit area of the galactic disk and the
subscript ``pc" on the emission measure indicates that it is measured in units
of cm$^{-6}$ pc.

In the absence of extinction, the \Ha\ photon intensity in Rayleighs is
\beq
I^*(\mha)=\frac{\alpha_{\mha}}{4\pi}\, EM
\;\left(\frac{1\,\R}{10^6/4\pi~{\rm photons}}\right)
     = 0.364 T_4^{-0.9}\empc~~~\R,
\label{eq:istar}
\eeq
where $\alpha_{\mha}\simeq 1.18\times 10^{-13}T_4^{-0.9}$ cm$^3$ s\e\ is the
effective recombination coefficient for \Ha\ \citep[from][]{O89}. The
relationship between the \Ha\ intensity normal to the disk and the surface
emissivity (eq. \ref{eq:ephas}) is almost independent of temperature,
\beq
I_\perp^*(\mha)=4.55\times 10^{-7}T_4^{-0.1}\ephas~~~\R.
\label{eq:iperpstar}
\eeq

These results can be readily extended to the recombination lines of helium,
such as He I $\lambda\, 5876$.  Ionization equilibrium in a column normal to
the disk gives
\beq
\zeta_{\Heo A}=y_{\Heo}\epheas\simeq \alt_\Heo
     \int dz\;n_e n({\rm He}^+),
\eeq
where $y_\Heo$ is the fraction of the helium ionizing photons that are
absorbed by neutral helium and $\alt_\Heo\simeq 2.73\times 10^{-13}
T_4^{-0.67}$ cm$^3$ s\e\ is the recombination coefficient to excited states of
He$^0$ \citep[from][]{O89}.  As written, the equation is approximate because,
in contrast to hydrogen, not all the recombinations to the ground state result
in the ionization of another helium atom.  The intensity of the $\lambda$ 5876
line is
\beq
I(5876)=\frac{\alpha_{5876}}{4\pi}\int dz\; n_e n({\rm He}^+),
\eeq
where $\alpha_{5876}\simeq 4.9\times 10^{-14}T_4^{-1.1}$ cm$^3$ s\e\ is the
effective recombination coefficient for this line \citep[from][]{O89}.  As a
result, we find
\beq
I^*(5876)=1.8\times 10^{-7} y_\Heo T_4^{-0.4}\epheas~~~\R.
\eeq

\subsection{Opacity in a Cloudy Medium}

The interstellar medium is highly inhomogeneous.  In the simplest
idealization, this inhomogeneity can be represented by clouds of opacity
$\tau_c$ embedded in a transparent intercloud medium.  The opacity in a cloudy
medium was derived by \citet{BF69}.  If the expected number of clouds along
the line of sight is $\nlos$, then the expected value of the ratio of the
observed intensity $I_\nu$ to the emitted intensity $I_{\nu 0}$ is
\beqa
I_\nu/I_{\nu 0} &=& \left(1+\nlos e^{-\tcn} + 
    \frac{1}{2!}\nlos^2 e^{-2\tcn}\dots
    \right) e^{-\nlos},\\
  & = & \exp[-\nlos(1-e^{-\tcn})],
\eeqa
where the different terms in the series represent the contribution to the
expected intensity if there are no clouds along the line of sight, one cloud,
two clouds, etc.  Thus, the expected value of the optical depth in a medium
composed of identical clouds is
\beq
\tau_\nu=\nlos\left(1-e^{-\tcn}\right).
\eeq
In the limit of small $\tcn$, the expected optical depth reduces to
$\nlos\tcn$, the value for a uniform medium.  On the other hand, in the limit
of large $\tcn$, the expected optical depth is $\nlos$, since the probability
that a ray will not encounter any clouds is $\exp(-\nlos)$.

If the total HI column density of the disk and halo is $\NHop$ and that of an
individual cloud is $\NHoc$, then
\beq
\ton=\frac{\NHop}{\NHoc}\left(1-e^{-\NHoc\sigma_\nu}\right),
\label{eq:ton}
\eeq
where $\sigma_\nu$ is the absorption cross section of the cloud
material.

If the clouds are not all identical, the above argument is
readily generalized \citep{BF69}.  The expected opacity is
\beq
\kappa_\nu=\frac{d\tau_\nu}{ds}=
	\frac{d\nlos}{ds}\, \left(1-e^{-\tcn}\right)
	\equiv\frac{1}{\lambda_c}\, \left(1-e^{-\tcn}\right),
\eeq
where $ds$ is an element of distance along the line of sight and $\lambda_c$
is the cloud mean free path.  In terms of the area of a cloud, $A_c$, and the
number of clouds per unit volume, $d\caln/dV$, the mean free path is
\beq
\frac{1}{\lambda_c}
 =A_c\frac{d\caln}{dV}.
\eeq
If the clouds have a spectrum of masses, then the expected opacity becomes
\beq
\kappa_\nu=\int dM\; A_c(M)\frac{d^2\caln}{dVdM}\,
    \left[1-e^{-\tcn(M)}\right].
\eeq

Finally, we note that the specific intensity calculated with this value of
$\kappa_\nu$ is approximately equal to that in the {\it intercloud} medium,
provided the filling factor of the intercloud medium $f_{ic}$ is close to
unity.  The solution of the equation of radiative transfer for a constant
source function is
\beq
I_\nu=S_\nu(1-e^{-\tau_\nu/\mu}).
\eeq
If $\tau_\nu$ or $\tcn$ is small, there is no distinction between the
intensity in the clouds and that in the intercloud medium, so we focus on the
case in which both optical depths are large.  In that case,
\beq
I_\nu\simeq S_\nu=\frac{j_\nu}{\kappa_\nu}
    \simeq j_\nu\lambda_c.
\eeq
In a cloudy medium, $j_\nu=j_{\nu,ic}f_{ic}$, where $j_{\nu,ic}$ is the
emissivity in the intercloud medium.  If the intercloud filling factor is
close to unity, then $I_\nu\simeq j_{\nu,ic}\lambda_c$, which is the expected
value of the intensity in the intercloud medium when $\tau_\nu$ and $\tcn$ are
large.

\clearpage

\clearpage

\begin{deluxetable}{ccccccc}
\tablewidth{0pc}
\tablecaption{Spectral Characteristics of SNR Spectra \label{tbl:phi}}
\tablehead{
\colhead{$n_a$} & \colhead{thermal} &
\colhead{$\phi_\Hy$} & \colhead{$\phi_{\Heo}$} &
\colhead{$\phi_{\Hep}$} & \colhead{$\phi_{X}$} &
\colhead{$Q$(He$^0$)/$Q$(H$^0$)} \\
\colhead{(cm$^{-3}$)} & \colhead{conduction} &
\colhead{} & \colhead{} &
\colhead{} & \colhead{} & \colhead{}
}
\startdata
0.04 & n & 0.32 & 0.096 & 0.038 & 0.0015 & 0.15 \\
0.04 & y & 0.31 & 0.091 & 0.039 & 0.0018 & 0.15 \\
0.1\tablenotemark{a}\phn & n & 0.39 &  0.11 & 0.052 & 0.0032 & 0.15 \\
0.1\phn & y & 0.38 &  0.12 & 0.054 & 0.0036 & 0.16 \\
0.3\phn & n & 0.42 &  0.14 & 0.076 & 0.0053 & 0.17 \\
0.3\phn & y & 0.42 &  0.13 & 0.077 & 0.0052 & 0.17 \\
1.0\phn & n & 0.39 &  0.16 & 0.10\phn & 0.0077 & 0.20 \\
1.0\phn & y & 0.36 &  0.14 & 0.10\phn & 0.0080 & 0.20 \\
\enddata
\tablenotetext{a}{Standard model}
\end{deluxetable}

\clearpage

\begin{deluxetable}{cccccc}
\tablewidth{0pc}
\tablecaption{Photoionization Due to SNR Emission \label{SNRphot}}
\tablehead{\colhead{$n_a$} & \colhead{thermal} & 
\colhead{$(\emppc/2)$\tablenotemark{a}} & 
\colhead{$h\bar\nu_\Hy$} & \colhead{$\calng(>\nu_\Hy)$} & 
\colhead{$\phi(>\nu_\Hy)$} \\
\colhead{(cm$^{-3}$)} & \colhead{conduction} & \colhead{(cm$^{-6}\,$pc)}
& \colhead{(eV)} & \colhead{$(10^{61}$~photons)} & \colhead{}}
\startdata
0.04 & n &  0.93 & 19.9 & 1.4 & 0.45 \\
0.04 & y &  0.92 & 19.9 & 1.4 & 0.45 \\
0.1\phn & n &  1.2 & 19.5 & 1.8 & 0.56 \\
0.1\phn & y &  1.2 & 19.8 & 1.8 & 0.56 \\
0.3\phn & n &  1.3 & 20.2 & 2.0 & 0.64 \\
0.3\phn & y &  1.3 & 20.1 & 2.0 & 0.64 \\
1.0\phn & n &  1.2 & 22.0 & 1.9 & 0.66 \\
1.0\phn & y &  1.2 & 21.4 & 1.8 & 0.62 \\
\enddata
\tablenotetext{a}{Emission measure for the half disk; assumes $T=8000$ K}
\end{deluxetable}

\clearpage

\begin{deluxetable}{lcccc}
\tablewidth{0pc}
\tablecaption{Comparison of Data with Model Results \label{tbl:lines}}
\tablehead{
\colhead{} & \multicolumn{2}{c}{Slow} &
\multicolumn{2}{c}{Fast} \\
\colhead{} & \colhead{Observed} & \colhead{Model\tablenotemark{a}} & 
\colhead{Observed} & \colhead{Model\tablenotemark{a}}
}
\startdata
$N$(\ion{S}{2})/(10$^{14}$ \sqc) & 12 & 12 & 8.8 & 8.8 \\
$N$(\ion{C}{2}$^*$)/(10$^{13}$ \sqc) & $8.4$ & 8.1 -- 8.5 &
2.3 & 2.5 -- 2.3 \\
$I({\rm H}\alpha)$ $(R)$ & 0.22 & 0.19 & 0.16 & 0.19 \\ \hline
$N$(\ion{H}{1})/(10$^{19}$ \sqc) & 6.8 & 7.6 -- 7.6 & 4.9 & 4.7 -- 4.3 \\
$N_e$/(10$^{18}$ \sqc) &  \nodata  & 4.9 -- 5.5 &  \nodata  & 16 -- 21 \\
$N$(\ion{S}{3})/$N$(\ion{S}{2}) & \nodata & 0.050 -- 0.051 & 0.10 & 
0.087 -- 0.10\phn \\
$I(5876)$ $(R)$ &  \nodata  & 0.0048 -- 0.0044 & \nodata & 
0.0053 -- 0.0051 \\
$I(6300)$ $(R)$ &  \nodata & 0.16 -- 0.53 &  \nodata  & 0.044 -- 0.12\phn \\
$I(6716)$ $(R)$ & 0.24 & 0.38 -- 0.92 & 0.19 & 0.14 -- 0.30 \\
$I(6584)$ $(R)$ & 0.22 & 0.058 -- 0.20\phn & 0.29 & 0.064 -- 0.19\phn \\
$\langle P \rangle$/(1000 cm$^{-3}$K) & \nodata & 10 -- 15 & 
\nodata & 1.5 -- 1.8 \\
$N$(\ion{H}{1})$_{X-ray}$/$N$(\ion{H}{1}) & \nodata & 1.028 -- 1.032 & 
\nodata & 1.172 -- 1.251 \\
\enddata
\tablenotetext{a}{The range of results is for models with $T=6000$ --
$8000$ K; see text.}
\end{deluxetable}

\end{document}